\documentclass[superscriptaddress,showpacs,fleqn,twocolumn]{revtex4}

\usepackage{amsmath}
\usepackage{graphicx}
\usepackage{dcolumn}
\usepackage[latin1]{inputenc}

\begin{document}
\title{Theoretical transition frequencies beyond 0.1 ppb accuracy in H$_2^+$, HD$^+$, and antiprotonic helium}

\author{Vladimir I. Korobov}
\affiliation{Bogoliubov Laboratory of Theoretical Physics, Joint Institute for Nuclear Research, Dubna 141980, Russia}

\author{Laurent Hilico}
\affiliation{Laboratoire Kastler Brossel, UPMC-Paris 6, ENS, CNRS; Case 74, 4 place Jussieu, 75005 Paris, France}
\affiliation{Universit\'e d'Evry-Val d'Essonne, Boulevard Fran\c{c}ois Mitterrand, 91025 Evry Cedex, France}

\author{Jean-Philippe Karr}
\affiliation{Laboratoire Kastler Brossel, UPMC-Paris 6, ENS, CNRS; Case 74, 4 place Jussieu, 75005 Paris, France}
\affiliation{Universit\'e d'Evry-Val d'Essonne, Boulevard Fran\c{c}ois Mitterrand, 91025 Evry Cedex, France}


\begin{abstract}
We present improved theoretical calculations of transition frequencies for the fundamental transitions $(L\!=\!0,v\!=\!1)\to(L'\!=\!0,v'\!=\!0)$ in the hydrogen molecular ions H$_2^+$ and HD$^+$ with a relative uncertainty $4\cdot10^{-11}$ and for the two-photon transitions in the antiprotonic helium atom with a relative uncertainty $10^{-10}$. To do that, the one-loop self-energy correction of order $\alpha(Z\alpha)^6$ is derived in the two Coulomb center approximation, and numerically evaluated in the case of the aforementioned transitions. The final results also include a complete set of other spin-independent corrections of order $m\alpha^7$. The leading order corrections of $\alpha^2\ln^3(Z\alpha)^{-2}(Z\alpha)^6$ are also considered that allows to estimate a magnitude of yet uncalculated contributions.
\end{abstract}

\maketitle

\section{Introduction}

The few-body bound-state quantum electrodynamics is a challenging problem. So far, a complete set of contributions up to order $m\alpha^6$ has been obtained and calculated for the two-electron helium-like atoms \cite{Pachucki10}, one-electron molecular ions \cite{korobov2008}, and antiprotonic helium \cite{korobov2008b}. A contribution of $m\alpha^7$ order including as well the nonlogarithmic part, has been obtained for the fine structure of helium $2^3P$ level in \cite{Pachucki09}. Four-particle systems were addressed in \cite{Komasa11} (H$_2$ and its isotopologues) and in \cite{YanDrake08} (lithium-like atoms). Recently, QED calculations up to $m\alpha^5$, and partially $m\alpha^6$ orders were carried out for the beryllium atom with four electrons \cite{Pachucki13}.

Progress in high-precision spectroscopy of three-body molecular or molecule-like systems has opened new possibilities for metrology of nucleus-to-electron mass ratios~\cite{roth2008}. One-photon ro-vibrational transitions were observed in HD$^+$ molecular ion with a relative uncertainty of 1-2 ppb~\cite{koelemeij2007,bressel2012}. Spectroscopy of two-photon transitions in antiprotonic helium at the 2-5 ppb level yielded a new value of the antiproton-to-electron mass ratio~\cite{hori2011}. These experiments, as well as others~\cite{karr2012,koelemeij2012}, are currently being developed towards higher precision, which motivates the evaluation of higher-order corrections in these systems. The importance of the $m_p/m_e$ problem is supported by recent experiments \cite{Rb11} with rubidium atoms, which allow to deduce a new value of the fine structure constant, $\alpha = e^2/(\hbar c)$, with a relative uncertainty $6.6\times10^{-10}$. Further improvement may be hindered by the present limits on the proton-to-electron mass ratio, which is determined by the latest CODATA adjustment \cite{codata2010} with a relative uncertainty  $4.1\times10^{-10}$ (see also \cite{verdu2004}).

Theoretical calculation of the complete set of QED corrections up to order $m\alpha^6$ has brought the theoretical uncertainty down to 0.3-0.4 ppb in H$_2^+$ or HD$^+$~\cite{korobov2008}, and about 1 ppb in antiprotonic helium~\cite{korobov2008b}. Very accurate leading order relativistic corrections are also available from \cite{Zhong09}. In the present work, we compute the complete set of $m\alpha^7$ order corrections including the one-loop self energy contribution, which represents the main source of theoretical uncertainty. This allows us to improve the accuracy by about one order of magnitude, thus making real the possibility of improving the knowledge of nucleus-to-electron mass ratios.

This paper is organized as follows: In Sec.~II the one-loop self-energy contribution at $m\alpha^7$ order for the hydrogen-like atoms is considered and a general formula for an arbitrary $(n,l)$ state, derived from comparison of \cite{czarnecki2005,jentschura2005} and \cite{pachucki1992,pachucki1993} results, is presented. In Sec.~III the low energy part is reconsidered to reformulate the result of Sec.~II in a form which is then suitable to be extended to the Coulomb two-center problem (Sec.~IV~A). A list of other contributions in $m\alpha^7$ and $m\alpha^8$ orders, which were also taken into account in the final results, are considered in Sec.~IV~B. Then examples of numerical calculations for the hydrogen isotope ions as well as for the antiprotonic helium are given in Sec.~V.

\section{The one-loop self-energy contribution at order $m\alpha^7$. Hydrogen-like case.}

As a starting point of our consideration we take the general result of Refs.~\cite{czarnecki2005,jentschura2005} for a bound electron in a field of external Coulomb potential, $V(\mathbf{r})=-Z/r$, written in the natural relativistic units ($\hbar=c=m=1$):
\begin{widetext}
\begin{equation}\label{JCP_eq}
\begin{array}{@{}l}
\displaystyle
\Delta E_{\rm se}^{(7)} =
   \frac{\alpha}{\pi}\Biggl\{(Z\alpha)^6\mathcal{L}_H+
   \left(
      \frac{5}{9}+\frac{2}{3}\ln\left[\frac{1}{2}(Z\alpha)^{-2}\right]
   \right)
   \left\langle
      4\pi\rho\>Q(E_0\!-\!H)^{-1}Q\,H_B
   \right\rangle
   +2\left\langle
      H_{so}\,Q(E_0\!-\!H)^{-1}Q\,H_B
   \right\rangle
\\[4mm]\displaystyle\hspace{12mm}
   +\left(
      \frac{779}{14400}+\frac{11}{120}\ln\left[\frac{1}{2}(Z\alpha)^{-2}\right]
   \right)\left\langle \boldsymbol{\nabla}^4V \right\rangle
   +\left(
      \frac{23}{576}+\frac{1}{24}\ln\left[\frac{1}{2}(Z\alpha)^{-2}\right]
   \right)
   \left\langle
      2\mathrm{i}\sigma^{ij}p^i\boldsymbol{\nabla}^2Vp^j
   \right\rangle
\\[4mm]\displaystyle\hspace{12mm}
   +\left(
      \frac{589}{720}+\frac{2}{3}\ln\left[\frac{1}{2}(Z\alpha)^{-2}\right]
   \right)\left\langle\left(\boldsymbol{\nabla}V\right)^2\right\rangle
   +\frac{3}{80}\left\langle 4\pi\rho\>\mathbf{p}^2 \right\rangle
   -\frac{1}{2}\left\langle\mathbf{p}^2H_{so}\right\rangle
   \Biggr\},
\end{array}
\end{equation}
\end{widetext}
where
\[
\begin{array}{@{}l}\displaystyle
H_B=-\frac{p^4}{8}+\frac{\pi}{2}\rho+H_{so},
\qquad
H_{so}=\frac{1}{4}\>\sigma^{ij}\nabla^iVp^j,
\\[4mm]\displaystyle
4\pi\rho\!=\!\Delta V,
\qquad
H_{so}^\delta = 2\mathrm{i}\sigma^{ij}p^i(\boldsymbol{\nabla}^2V)p^j.
\end{array}
\]
and $\sigma^{ij}=[\sigma^i\sigma^j]/(2i)=\epsilon^{ijk}\sigma^k$.
Brackets denote averaging on the nonrelativistic bound state wave function $\psi_0$, $E_0$ and
$
H = p^2/2 + V
$
are, respectively, the nonrelativistic energy of the state and the nonrelativistic Hamiltonian. Here $Q$ is a projector operator on the subspace orthogonal to $\psi_0$. $\mathcal{L}_H$ is the low-energy photon contribution or the relativistic Bethe logarithm for the hydrogen atom state (assuming $Z=1$). The above result is valid {\em for all states with nonzero angular momentum and for the normalized difference of $S$ states},
\[
\Delta_n=n^3\Delta E(nS)-\Delta E(1s).
\]

A more general expression which would also be valid for individual $S$ states, {\em will differ from (\ref{JCP_eq}) only by a term proportional to the delta function, $\delta(\mathbf{r})$}. To get the unknown contribution for the hydrogen case, we have to compare with the result of \cite{pachucki1992,pachucki1993} obtained for the $1S$ state of hydrogen. For this purpose, the expectation values in (\ref{JCP_eq}) which are divergent for individual S states should first be regularized; we will use a regularization by cut-off of a small $r$ spherical domain around the nucleus. Any two such regularizations differ by a term proportional to the delta function, so that the result will still differ from an expression valid for all states by a delta-function term.

To that end let us introduce two functionals $\mathcal{Q}$ and $\mathcal{R}$

\begin{widetext}
\begin{equation}\label{Q}
\begin{array}{@{}l}
\displaystyle
\mathcal{Q} =
   \lim_{r_0\to0}
   \left\{
   \left\langle
      \frac{1}{4\pi r^3}
   \right\rangle_{\!\!r_0}\!
   +\left(\ln{r_0}\!+\!\ln\alpha\!+\!\gamma_E\right)\left\langle\delta(\mathbf{r})\right\rangle
   \right\}
\\[3mm]\displaystyle\hspace{30mm}
 = -\frac{(Z\alpha)^3}{\pi n^3}
   \left[
      -\frac{1}{2}\ln{Z^{-2}}+\psi(n)-\psi(1)-\ln{\frac{n}{2}}
      -\frac{1}{2}+\frac{1}{2n}
   \right],\end{array}
\end{equation}
\vspace{-3mm}
\begin{equation}\label{R}
\begin{array}{@{}l}
\displaystyle
\mathcal{R} =
   \lim_{r_0\to0}
   \left\{
   \left\langle
      \frac{1}{4\pi r^4}
   \right\rangle_{\!\!r_0}\!
   -\left[
      \frac{1}{r_0}\left\langle\delta(\mathbf{r})\right\rangle
      +\left(\ln{r_0}\!+\!\ln\alpha\!+\!\gamma_E\right)\left\langle\delta'(\mathbf{r})\right\rangle
   \right]
   \right\}
\\[3mm]\displaystyle\hspace{30mm}
   =
   \frac{2(Z\alpha)^4}{\pi n^3}
   \left[
      -\frac{1}{2}\ln{Z^{-2}}+\psi(n)-\psi(1)-\ln{\frac{n}{2}}
      -\frac{5}{3}+\frac{1}{2n}+\frac{1}{6n^2}
   \right].
\end{array}
\end{equation}
where
\[
\begin{array}{@{}l}
\displaystyle
\left\langle\phi_1|\delta'(\mathbf{r})|\phi_2\right\rangle =
\left\langle\phi_1\left|
   \frac{\mathbf{r}}{r}\boldsymbol{\nabla}\delta(\mathbf{r})
\right|\phi_2\right\rangle
=   -\left\langle\partial_r\phi_1|\delta(\mathbf{r})|\phi_2\right\rangle
   -\left\langle\phi_1|\delta(\mathbf{r})|\partial_r\phi_2\right\rangle,
\end{array}
\]
$\left\langle \>\> \right\rangle_{r_0}$ denotes integration outside a sphere of radius $r_0$. The last line in Eqs.~(\ref{Q}-\ref{R}) contains an expectation value of $\mathcal{Q}$ (or $\mathcal{R}$) for $nS$ states of hydrogen-like atoms. Using these expressions all divergent matrix elements appearing in Eq.~(\ref{JCP_eq}) may be redefined in a finite form: 
\begin{subequations} \label{finite}
\begin{equation}
\left\langle 4\pi\rho\>\mathbf{p}^2 \right\rangle_{\rm fin} =
   8\pi (Z\alpha)^2\mathcal{R}
   +16\pi (Z\alpha)^3\mathcal{Q}
   +4E_0\left\langle V^2 \right\rangle
   -2\left\langle\mathbf{p}V^2\mathbf{p}\right\rangle
   +2E_0\left\langle 4\pi\rho \right\rangle
\end{equation}
\vspace*{-3mm}
\begin{equation}
\left\langle \left[\boldsymbol{\nabla}^4V\right] \right\rangle_{\rm fin} =
   -16\pi (Z\alpha)^2\mathcal{R}
   -32\pi (Z\alpha)^3\mathcal{Q}
   -8E_0\left\langle V^2 \right\rangle
   +4\left\langle\mathbf{p}V^2\mathbf{p}\right\rangle
   +2\left\langle \mathbf{p}(4\pi\rho)\mathbf{p} \right\rangle
   -4E_0\left\langle 4\pi\rho \right\rangle
\end{equation}
\vspace*{-3mm}
\begin{equation} \label{order2}
\begin{array}{@{}l}
\displaystyle
\left\langle
   4\pi\rho\>Q(E\!-\!H)^{-1}Q\,H_B
\right\rangle_{\rm fin} =
\left\langle
   H'^{(1)}\>Q(E\!-\!H)^{-1}Q\,H'^{(2)}
\right\rangle
\\[3mm]\hspace{20mm}\displaystyle
+\frac{1}{4}
\Bigl[
   4\pi (Z\alpha)^2\mathcal{R}
   +16\pi (Z\alpha)^3\mathcal{Q}
   +8E_0\left\langle V^2 \right\rangle
   -4E_0^2\left\langle V \right\rangle
   +\left\langle H^{(1)}\right\rangle \langle V \rangle
   -8\left\langle H^{(2)}\right\rangle \langle V \rangle
\Bigr]
\end{array}
\end{equation}
\end{subequations}
\end{widetext}
In the last expression $H^{(1)} = 4\pi\rho$, $H^{(2)} = H_B$, which are transformed \cite{pachucki2000,korobov2007} as
\[
\left\{
\begin{array}{l}
\displaystyle
H'^{(1)}=-(E_0-H_0)U_1-U_1(E_0-H_0)+H^{(1)}
\\[1mm]\displaystyle
H'^{(2)}=-(E_0-H_0)U_2-U_2(E_0-H_0)+H^{(2)}
\end{array}
\right.
\]
to eliminate the divergent part from the second order term, here $U_1 = 2V$ and $U_2 = -\frac{1}{4}V$.

Thus obtained expression should be compared with the complete result for a $1s$ state \cite{pachucki1992,pachucki1993}:
\begin{widetext}
\begin{equation}
\begin{array}{@{}l}\displaystyle
\Delta E_{\rm se}^{(7)}(1S) =
   \frac{\alpha(Z\alpha)^6}{\pi}
   \biggl\{
      -\ln^2\bigl[(Z\alpha)^{-2}\bigr]
      +\left[\frac{28}{3}\ln2-\frac{21}{20}\right]
      \ln\bigl[(Z\alpha)^{-2}\bigr]
      -30.92414946(1)
   \biggr\}
\end{array}
\end{equation}
which yields (using $\mathcal{L}(1S) = -27.25990948(1)$~\cite{jentschura2005})
\begin{equation}\label{3}
\begin{array}{@{}l}
\displaystyle
\Delta E_{\rm se}^{(7)} =
   \frac{\alpha}{\pi}\Biggl\{(Z\alpha)^6\mathcal{L}_H+
   \left(
      \frac{5}{9}+\frac{2}{3}\ln\left[\frac{1}{2}(Z\alpha)^{-2}\right]
   \right)
   \left\langle
      4\pi\rho\>Q(E\!-\!H)^{-1}Q\,H_B
   \right\rangle_{\rm fin}
   +2\left\langle
      H_{so}\,Q(E\!-\!H)^{-1}Q\,H_B
   \right\rangle
\\[3mm]\displaystyle\hspace{18mm}
   +\left(
      \frac{779}{14400}+\frac{11}{120}\ln\left[\frac{1}{2}(Z\alpha)^{-2}\right]
   \right)\left\langle \boldsymbol{\nabla}^4V \right\rangle_{\rm fin}
   +\left(
      \frac{23}{576}+\frac{1}{24}\ln\left[\frac{1}{2}(Z\alpha)^{-2}\right]
   \right)
   \left\langle
      H_{so}^\delta
   \right\rangle
\\[3mm]\displaystyle\hspace{18mm}
   +\left(
      \frac{589}{720}+\frac{2}{3}\ln\left[\frac{1}{2}(Z\alpha)^{-2}\right]
   \right)\left\langle\left(\boldsymbol{\nabla}V\right)^2\right\rangle_{\rm fin}
   +\frac{3}{80}\left\langle 4\pi\rho\>\mathbf{p}^2 \right\rangle_{\rm fin}
   -\frac{1}{2}\left\langle\mathbf{p}^2H_{so}\right\rangle
\\[3mm]\displaystyle\hspace{18mm}
   +\biggl[
      -\ln^2\bigl(\alpha^{-2}\bigr)
      +\left(\frac{16}{3}\ln{2}-\frac{1}{4}\right)\ln\bigl(\alpha^{-2}\bigr)
\\[2mm]\displaystyle\hspace{23mm}
      +\ln^2{Z^{-2}}
      +\left(\frac{10}{3}\ln{2}+\frac{37}{15}\right)\ln{Z^{-2}}
      -0.81971202(1)
   \biggr]
   (Z\alpha)^2\left\langle \pi\rho \right\rangle
   \Biggr\}
\end{array}
\end{equation}
\end{widetext}
for the hydrogen-like atom.

\section{The low-energy part: Redefining the relativistic Bethe logarithm to atomic units}

From this point and in what follows we will use atomic units: $m_e=\hbar=e=1$.

In expressions (1) and (6), the relativistic Bethe logarithm $\mathcal{L}_H$ is defined using the energy scale $Z^2 E_h$, which is well suited for the hydrogenic case, but becomes irrelevant for a system with two Coulomb centers of charges $Z_1$, $Z_2$. For this reason, we have to redefine the relativistic Bethe logarithm $\mathcal{L}(Z,n,l)$ in a.u.

The low-energy part has been considered in more details in \cite{korobov2013}. Here we will try to elucidate only the key points of the derivation.

The relativistic Bethe logarithm is determined in integral form as follows
\begin{equation}\label{BLau}
\mathcal{L}(Z,n,l) = \frac{2}{3}\int_0^{E_h} kdk P_{\alpha^2}^{(1)}(k)
   + \frac{2}{3}\int_{E_h}^\infty kdk P_{\alpha^2}^{(2)}(k),
\end{equation}
where $E_h$ is the Hartree energy.

\begin{widetext}
The integrand is a function of energy and is a sum of various contributions:

a) relativistic corrections to the wave function
\begin{equation}\label{Prc1}
\begin{array}{@{}l}
\displaystyle
P_{rc}^{(1)}(k) =
   2\left\langle
      H_BQ(E_0-H)^{-1}Q\mathbf{p}\left(E_0-H-k\right)^{-1}\mathbf{p}
   \right\rangle
\\[2mm]\hspace{35mm}\displaystyle
   +\left\langle
      \mathbf{p}\left(E_0-H-k\right)^{-1}
      \Bigl(H_B-\left\langle H_B \right\rangle\Bigr)
      \left(E_0-H-k\right)^{-1}\mathbf{p}
   \right\rangle;
\end{array}
\end{equation}

b) modification of the vertex interactions
\begin{equation} \label{Prc2}
P_{rc}^{(2)}(k) =
   \left\langle
      \left(-p^2p^i-\frac{1}{2}\sigma^{ij}\nabla^jV\right)
      \left(E_0-H-k\right)^{-1}p^i
   \right\rangle;
\end{equation}

c) nonrelativistic quadrupole contribution
\begin{equation} \label{Pnq}
\begin{array}{@{}l}
\displaystyle
P_{nq}(k) =
   \frac{3k^2}{8\pi}\int_S d\Omega_\mathbf{n}
      \left(\delta^{ij}\!-\!n^in^j\right)
      \biggl\{
      \left\langle
         p^i(\mathbf{n\cdot r})\left(E_0\!-\!H\!-\!k\right)^{-1}(\mathbf{n\cdot r})p^i
      \right\rangle
      -\left\langle
         p^i(\mathbf{n\cdot r})^2\left(E_0\!-\!H\!-\!k\right)^{-1}p^i
      \right\rangle
      \biggr\},
\end{array}
\end{equation}
where $\mathbf{k} = k\mathbf{n}$.

The complete contribution is $P_{\alpha^2}(k) = P_{rc}^{(1)}(k)+P_{rc}^{(2)}(k)+P_{nq}(k)$.
Its asymptotic expansion for large $k$ may be written in operator form up to terms of $\mathcal{O}(1/k^2)$ (see Appendix for asymptotic expansion of separate contributions):
\begin{equation}\label{asymp}
\begin{array}{@{}l}\displaystyle
P_{\alpha^2}(k)
 = -\frac{1}{2}
      \bigl\langle
         \boldsymbol{\nabla}^2
      \bigr\rangle
   +\frac{2}{k}\left\langle
      \left( H_B\!-\!\left\langle H_B \right\rangle\right)
      (E_0\!-\!H)^{-1}\boldsymbol{\nabla}^2
   \right\rangle
   +\frac{4}{5k}
      \bigl\langle
         \boldsymbol{\nabla}^4
      \bigr\rangle
   -\frac{1}{2k}\left\langle (\boldsymbol{\nabla}^2V) \right\rangle
\\[3mm]\displaystyle\hspace{15mm}
   +\frac{\sqrt{2}}{k^{3/2}}\,\pi Z^2\left\langle\delta(\mathbf{r})\right\rangle
   -\frac{3\ln{k}}{k^2}\,\pi Z^3\left\langle\delta(\mathbf{r})\right\rangle
   +\frac{1}{k^2}\left(5\ln{2}+\frac{37}{10}\right)\>\pi Z^3\left\langle \delta(\mathbf{r}) \right\rangle
\\[3mm]\displaystyle\hspace{15mm}
   +\frac{1}{k^2}\left\langle
      \left( H_B\!-\!\left\langle H_B \right\rangle\right)
      (E_0\!-\!H)^{-1}(\boldsymbol{\nabla}^2V)
   \right\rangle_{\rm fin}
   +\frac{1}{k^2}\left\langle(\boldsymbol{\nabla}V)^2\right\rangle_{\rm fin}
   +\frac{11}{80k^2}\left\langle(\boldsymbol{\nabla}^4V)\right\rangle_{\rm fin}
   +\frac{1}{16k^2}
   \left\langle
      H_{so}^\delta
   \right\rangle
   +\dots
\end{array}
\end{equation}
\end{widetext}

The finite expectation values are defined in a similar way as in the previous section, taking into account that the functionals $\mathcal{Q}$ and $\mathcal{R}$ should be accordingly modified:
\begin{equation}
\mathcal{Q} =
   \lim_{r_0\to0}
   \left\{
   \left\langle
      \frac{1}{4\pi r^3}
   \right\rangle_{\!\!r_0}\!
   +\left(\ln{r_0}\!+\!\gamma_E\right)\left\langle\delta(\mathbf{r})\right\rangle
   \right\},
\end{equation}
\vspace{-3mm}
\begin{equation}
\begin{array}{@{}l}
\displaystyle
\mathcal{R} =
   \lim_{r_0\to0}
   \Biggl\{
   \left\langle
      \frac{1}{4\pi r^4}
   \right\rangle_{\!\!r_0}\!
   -\biggl[
      \frac{1}{r_0}\left\langle\delta(\mathbf{r})\right\rangle
\\[3mm]\displaystyle\hspace{30mm}
      +\left(\ln{r_0}\!+\!\gamma_E\right)\left\langle\delta'(\mathbf{r})\right\rangle
   \biggr]
   \Biggr\}.
\end{array}
\end{equation}

As is discussed in \cite{jentschura2005,korobov2013} we have to subtract the leading terms of expansion (\ref{asymp}):
\begin{subequations}\label{Pint}
\begin{equation}
P_{\alpha^2}^{(1)}(k) = P_{\alpha^2}(k)-F_{\alpha^2}-\frac{A_{\alpha^2}}{k}-\frac{B_{\alpha^2}}{k^{3/2}}
\end{equation}
and
\begin{equation}
\begin{array}{@{}l}\displaystyle
P_{\alpha^2}^{(2)}(k) = P_{\alpha^2}(k)-F_{\alpha^2}-\frac{A_{\alpha^2}}{k}-\frac{B_{\alpha^2}}{k^{3/2}}
\\[3mm]\displaystyle\hspace{30mm}
   -\frac{C_{\alpha^2}\ln{k}}{k^2}-\frac{D_{\alpha^2}}{k^2}.
\end{array}
\end{equation}
\end{subequations}
Constants $F$, $A$, $B$, $C$, and $D$ are taken by evaluating expectation values of operators appearing in the expansion (\ref{asymp}) for the nonrelativistic wave function of a particular state.

The previous definition of the relativistic Bethe logarithm $\mathcal{L}_H$ assumes scaling to $(Z\alpha)=1$, and thus it may be expressed in atomic units as
\begin{equation}\label{BL0}
\begin{array}{@{}l}\displaystyle
\mathcal{L}_H(n,l) = Z^{-6}\Bigl[\frac{2}{3}\int_0^{Z^2E_h} kdk P_{\alpha^2}^{(1)}(k)
\\[4mm]\hspace{30mm}\displaystyle
   + \frac{2}{3}\int_{Z^2E_h}^\infty kdk P_{\alpha^2}^{(2)}(k)\Bigr].
\end{array}
\end{equation}

\begin{figure*}[t]
\begin{center}
\includegraphics[width=0.4\textwidth]{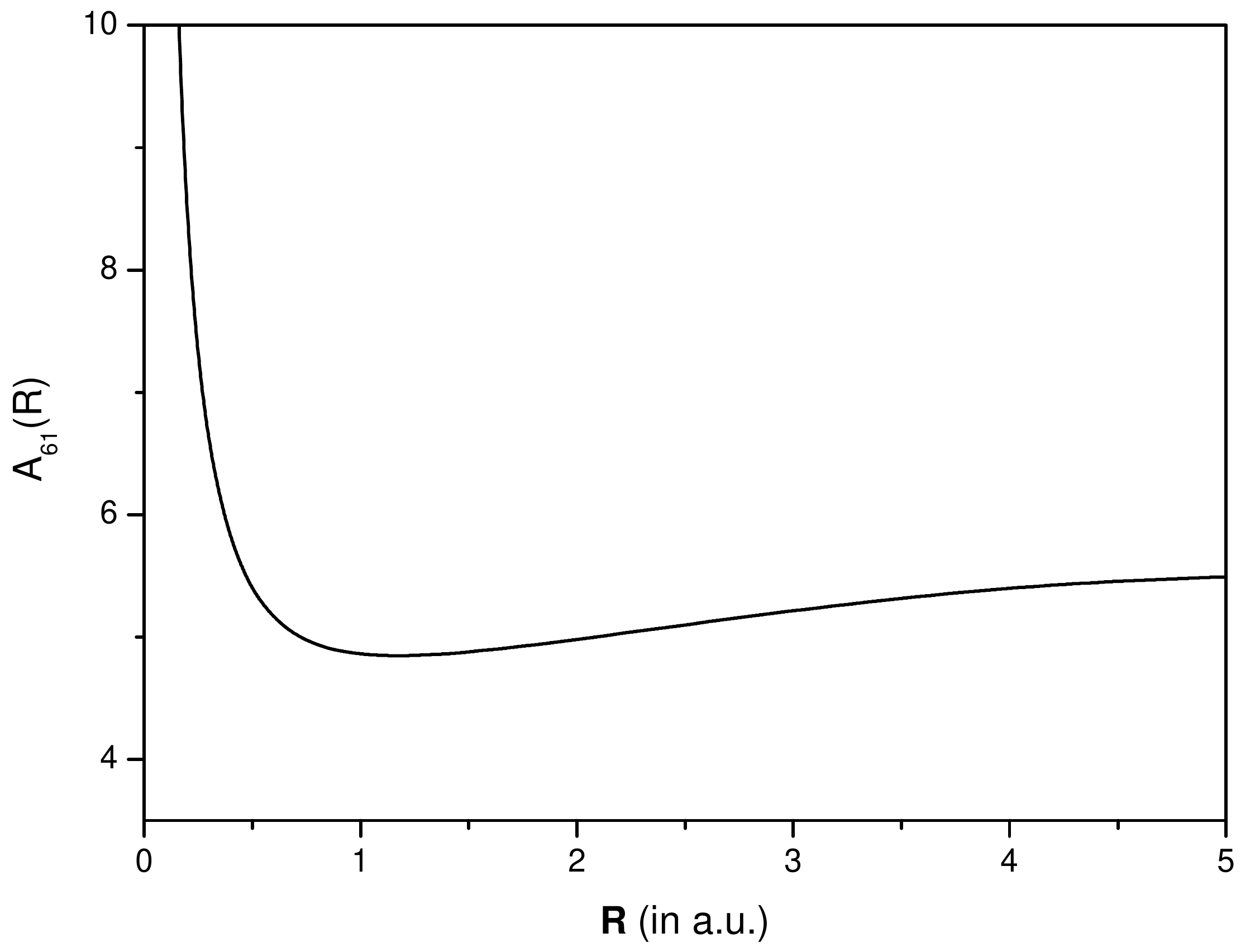}\hspace{4mm}
\includegraphics[width=0.4\textwidth]{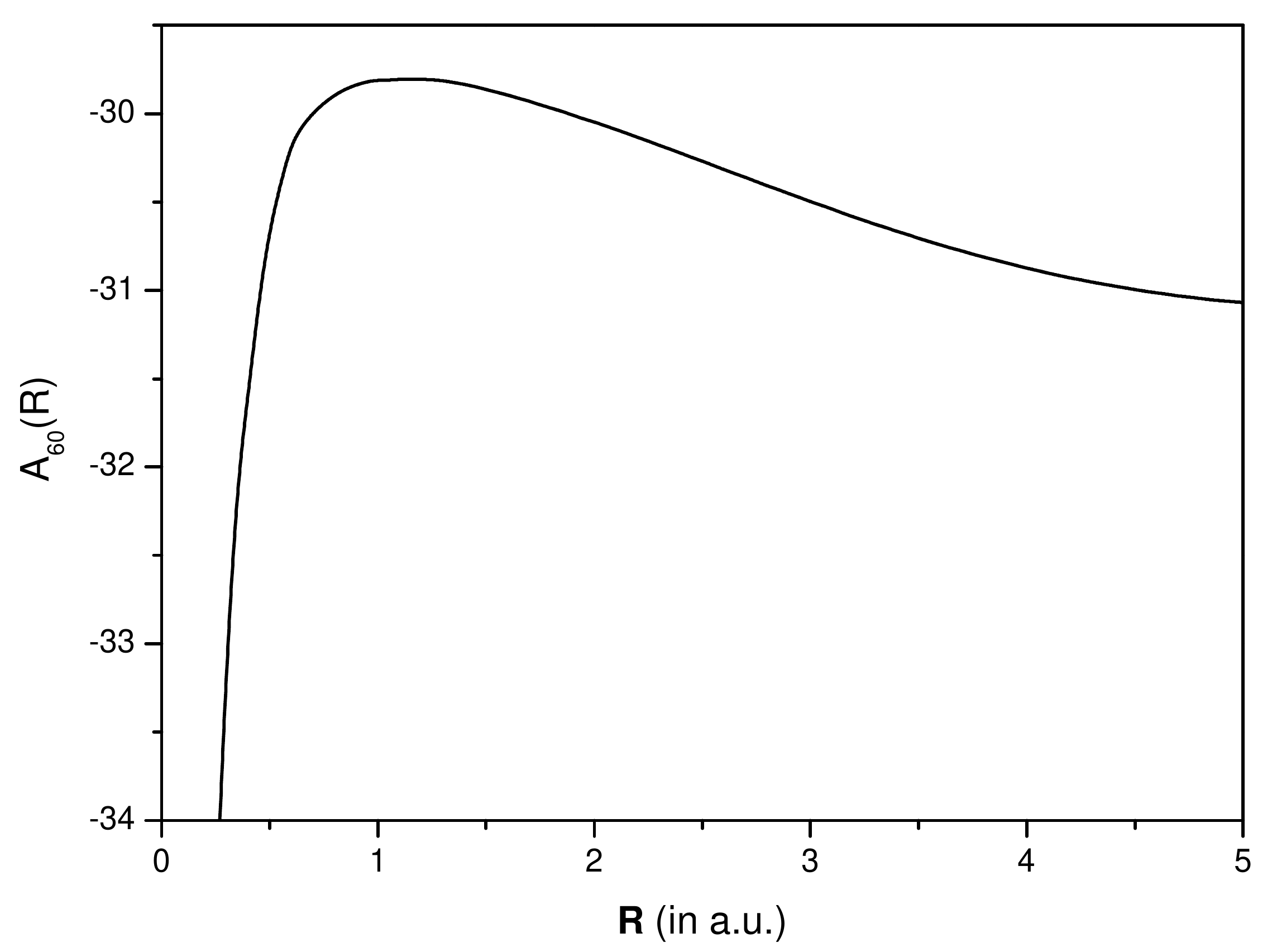}
\end{center}
\caption{The coefficients $A_{61}(R)$ and $A_{60}(R)$ for the ground ($1s\sigma$) electronic state of the two-center problem ($Z_1 = Z_2 = 1$, H$_2^+$ case) as a function of a bond length $R$.}\label{A6}
\end{figure*}

Comparing (\ref{BL0}) with (\ref{BLau}) one gets a relation between the two definitions of the relativistic Bethe logarithm:
\begin{widetext}
\begin{equation}
\begin{array}{@{}l}\displaystyle
\mathcal{L}(Z,n,l) = Z^6\mathcal{L}_H(n,l)
   +\frac{2}{3}\int_{E_h}^{Z^2E_h} dk
    \Biggl\{
    \left[
       \frac{3\ln{k}}{k} - \frac{1}{k}\left(5\ln{2}+\frac{37}{10}\right)
    \right]
    \pi Z^3\left\langle\delta(\mathbf{r})\right\rangle
\\[3mm]\displaystyle\hspace{12mm}
   -\frac{1}{k}
   \left[
   \left\langle
      \left( H_B\!-\!\left\langle H_B \right\rangle\right)
      (E_0\!-\!H)^{-1}(\boldsymbol{\nabla}^2V)
   \right\rangle_{\rm fin}
   +\left\langle(\boldsymbol{\nabla}V)^2\right\rangle_{\rm fin}
   +\frac{11}{80}\left\langle(\boldsymbol{\nabla}^4V)\right\rangle_{\rm fin}
   +\frac{1}{16}
   \left\langle
      H_{so}^\delta
   \right\rangle
   \right]
   \Biggr\}
\\[4mm]\hspace{8mm}\displaystyle
 = Z^6\mathcal{L}_H(n,l)
   +\frac{Z^6}{n^3}\left[
      \ln^2{Z^{-2}}
      +\ln{Z^{-2}}\left(\frac{10}{3}\ln{2}\!+\!\frac{37}{15}\right)
   \right]
\\[3mm]\displaystyle\hspace{12mm}
   +\ln{Z^{-2}}
   \left[
   \frac{2}{3}\left\langle
      \left( H_B\!-\!\left\langle H_B \right\rangle\right)
      (E_0\!-\!H)^{-1}(\boldsymbol{\nabla}^2V)
   \right\rangle_{\rm fin}
   +\frac{2}{3}\left\langle(\boldsymbol{\nabla}V)^2\right\rangle_{\rm fin}
   +\frac{11}{120}\left\langle(\boldsymbol{\nabla}^4V)\right\rangle_{\rm fin}
   +\frac{1}{24}
   \left\langle
      H_{so}^\delta
   \right\rangle
   \right]\hspace*{-25mm}
\end{array}
\end{equation}
and now substituting this into expression (\ref{3}) we immediately get the general expression for the one-loop self-energy correction in the $m\alpha^7$ order in atomic units:
\begin{equation}\label{a7au}
\begin{array}{@{}l}
\displaystyle
\Delta E_{\rm se}^{(7)} =
   \frac{\alpha^5}{\pi}\Biggl\{\mathcal{L}(Z,n,l)+
   \left(
      \frac{5}{9}+\frac{2}{3}\ln\left[\frac{\alpha^{-2}}{2}\right]
   \right)
   \left\langle
      4\pi\rho\>Q(E\!-\!H)^{-1}Q\,H_B
   \right\rangle_{\rm fin}
\\[4mm]\displaystyle\hspace{14mm}
   +2\left\langle
      H_{so}\,Q(E\!-\!H)^{-1}Q\,H_B
   \right\rangle
   +\left(
      \frac{779}{14400}+\frac{11}{120}\ln\left[\frac{\alpha^{-2}}{2}\right]
   \right)\left\langle \boldsymbol{\nabla}^4V \right\rangle_{\rm fin}
\\[4mm]\displaystyle\hspace{14mm}
   +\left(
      \frac{23}{576}+\frac{1}{24}\ln\left[\frac{\alpha^{-2}}{2}\right]
   \right)
   \left\langle
      H_{so}^\delta
   \right\rangle
   +\left(
      \frac{589}{720}+\frac{2}{3}\ln\left[\frac{\alpha^{-2}}{2}\right]
   \right)\left\langle\left(\boldsymbol{\nabla}V\right)^2\right\rangle_{\rm fin}
   +\frac{3}{80}\left\langle 4\pi\rho\>\mathbf{p}^2 \right\rangle_{\rm fin}
   -\frac{1}{2}\left\langle\mathbf{p}^2H_{so}\right\rangle
\\[4mm]\displaystyle\hspace{14mm}
   +Z^2\biggl[
   -\ln^2\bigl[\alpha^{-2}\bigr]
   +\left[\frac{16}{3}\ln2-\frac{1}{4}\right]\ln\bigl[\alpha^{-2}\bigr]
   -0.81971202(1)
   \biggr]\left\langle \pi\rho \right\rangle
   \Biggr\}
\end{array}
\end{equation}
\end{widetext}
This formula is quite general and may be extended to the case of external electric field of two (or more) Coulomb sources. One may check that the above expression matches the result of Erickson and Yennie for the logarithmic term for an arbitrary $nS$ state of the hydrogen atom~\cite{erickson1965}.

\begin{figure*}[t]
\begin{center}
\includegraphics[width=0.4\textwidth]{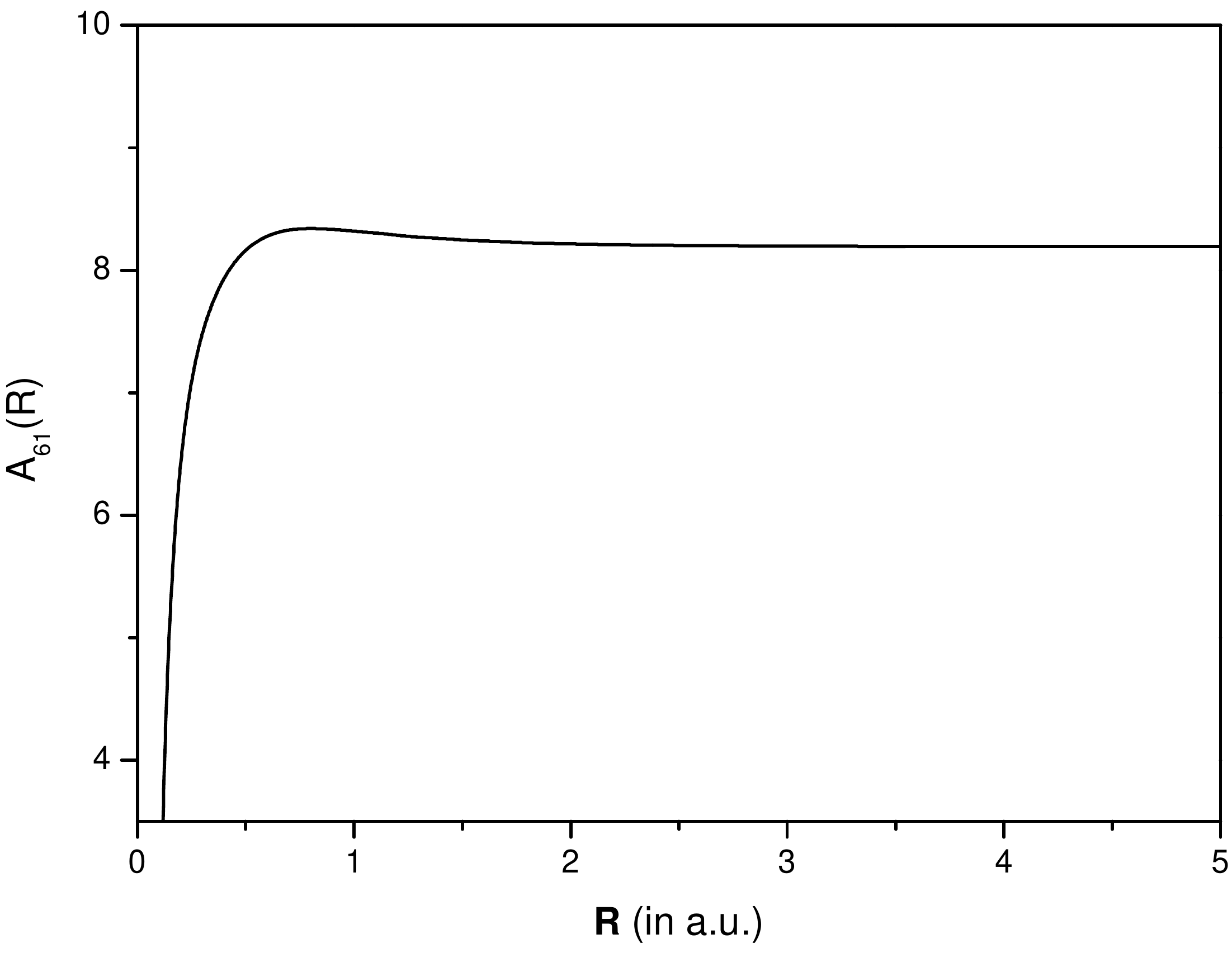}\hspace{4mm}
\includegraphics[width=0.4\textwidth]{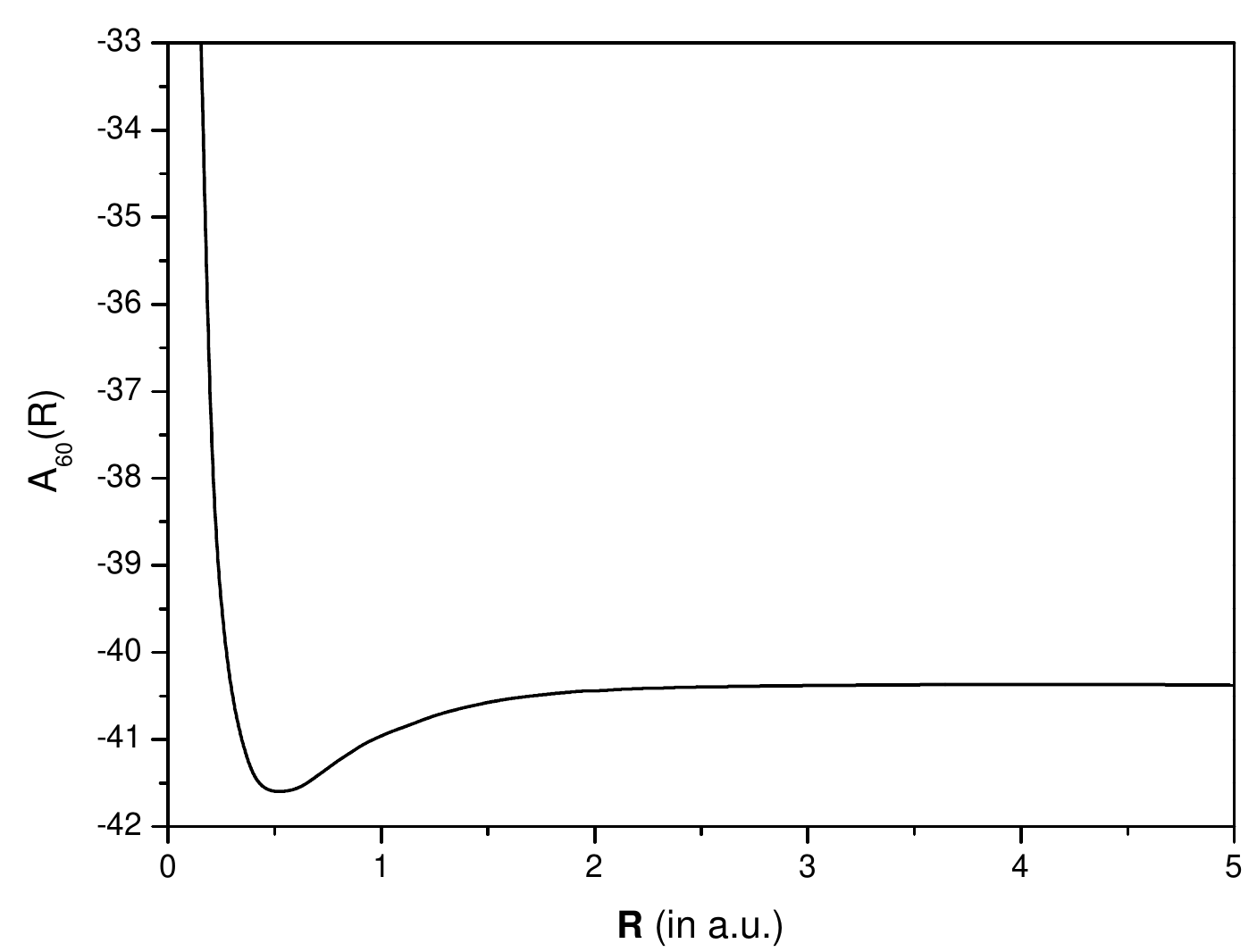}
\end{center}
\caption{The coefficients $A_{61}(R)$ and $A_{60}(R)$ for the ground ($1s\sigma$) electronic state of the two-center problem ($Z_1 = 2$, $Z_2 = -1$, He$^+\bar{p}\,$ case) as a function of a bond length $R$.}\label{A6_pbarHe}
\end{figure*}

\section{Coulomb two-center problem}

\subsection{One-loop self energy}

For the case of the two-center Coulomb problem one needs to replace the delta-function distribution, $Z^2 \langle \pi\rho \rangle$, in the last line of Eq.~(\ref{a7au}) by a distribution:
\begin{equation}
V_{\delta} =
   \pi \Bigl[
      Z_1^3 \delta(\mathbf{r}_1) + Z_2^3 \delta(\mathbf{r}_2)
   \Bigr].
\end{equation}

To present our results we will adopt a similar notation as for hydrogen-like ions~\cite{sapirstein1990}:
\begin{equation}\label{E_expan}
\Delta E_{\rm se}^{(7)} = \frac{\alpha^5}{\pi} \left\langle V_{\delta} \right\rangle
      \Bigl[
         A_{62} \ln^2[\alpha^{-2}]
         +A_{61} \ln[\alpha^{-2}] + A_{60}
      \Bigr],
\end{equation}
where $A_{62} = -1$; expressions for $A_{61}$ and $A_{60}$ coefficients are obtained by comparison between Eqs.~(\ref{a7au}) and~(\ref{E_expan}).
\begin{widetext}
\begin{equation}\label{A61_A60}
\begin{array}{@{}l}\displaystyle
A_{61}(R) =
\left[
   \frac{2}{3}
   \left\langle
      4\pi\rho\>Q(E\!-\!H)^{-1}Q\,H_B
   \right\rangle_{\rm fin}
   +\frac{11}{120}
   \left\langle \boldsymbol{\nabla}^4V \right\rangle_{\rm fin}
   +\frac{2}{3}\left\langle\left(\boldsymbol{\nabla}V\right)^2\right\rangle_{\rm fin}
   +\left(\frac{16}{3}\ln{2}\!-\!\frac{1}{4}\right)
   \left\langle V_{\delta} \right\rangle
\right]/\left\langle V_{\delta} \right\rangle
\\[3mm]\displaystyle
A_{60}(R) =
\biggl[
   \left(\frac{5}{9}-\frac{2}{3}\ln{2}\right)
   \left\langle
      4\pi\rho\>Q(E\!-\!H)^{-1}Q\,H_B
   \right\rangle_{\rm fin}
   +\left(
      \frac{779}{14400}-\frac{11}{120}\ln{2}
   \right)
   \left\langle \boldsymbol{\nabla}^4V \right\rangle_{\rm fin}
\\[3mm]\hspace{30mm}\displaystyle
   +\left(
      \frac{589}{720}-\frac{2}{3}\ln{2}
   \right)
   \left\langle\left(\boldsymbol{\nabla}V\right)^2\right\rangle_{\rm fin}
   +\frac{3}{80}\left\langle 4\pi\rho\>\mathbf{p}^2 \right\rangle_{\rm fin}
   -0.81971202(1)\left\langle V_{\delta} \right\rangle
   +\mathcal{L}(R)
\biggr]/\left\langle V_{\delta} \right\rangle.
\end{array}
\end{equation}
\end{widetext}
Since we are interested in the spin-independent part of transition frequency we have dropped out the terms from Eq.~(\ref{a7au}), which correspond to the spin-orbit interaction. They will be considered elsewhere.

The coefficients $A_{61}$ and $A_{60}$ now may be calculated by averaging of the "effective" potentials over the vibrational wave function of a three-body state (see Sec.~V and Figs.~\ref{A6}--\ref{A6_pbarHe}).

\begin{table}[t]
\begin{center}
\begin{tabular}{@{\hspace{2mm}}l@{\hspace{7mm}}d@{\hspace{7mm}}d@{\hspace{-2mm}}}
\hline\hline
\vrule height 10.5pt width 0pt depth 3.5pt
 & \mbox{H}_2^+ & \mbox{HD}^+ \\
\hline
\vrule height 10pt width 0pt
$\Delta E_{nr}$ & 65\,687\,511.0714   & 57\,349\,439.9733   \\
$\Delta E_{\alpha^4}$ &   1091.0397   &          958.1510   \\
$\Delta E_{\alpha^5}$ &   -276.5450   &         -242.1263   \\
$\Delta E_{\alpha^6}$ &     -1.9969   &           -1.7481   \\
$\Delta E_{\alpha^7}$ &      0.138(2) &            0.120(2) \\
$\Delta E_{\alpha^8}$ &      0.001(1) &            0.001(1) \\
\hline
\vrule height 10pt width 0pt depth 3.5pt
$\Delta E_{tot}$& 65\,688\,323.708(2) & 57\,350\,154.371(2) \\
\hline\hline
\end{tabular}
\end{center}
\caption{Summary of contributions to the $(v=0,L=0) \to (v'=1,L'=0)$ fundamental transition frequency of H$_2^+$ and HD$^+$ molecular ions (in MHz).}\label{h2p-hdp}
\end{table}

\subsection{Other contributions}

In addition to the one-loop self-energy correction, we computed several other contributions at orders $m\alpha^7$ and $m\alpha^8$, which did not require extensive calculations. Using the results from~\cite{eides} we see that most of the terms are proportional to $|\Psi(0)|^2$.

To better identify the most relevant terms, we give numerical values of all the correction terms to the fundamental vibrational transition frequency  $(v=0,L=0) \to (v=1,L=0)$ in H$_2^+$ (see Sec V for details on the numerical calculations). For comparison, the one-loop self energy term we have just obtained gives a contribution:
\begin{equation}
\Delta E_{se}^{(7)} \approx 125\pm2\mbox{ kHz}.
\end{equation}
The uncertainty here is primarily due to numerical inaccuracy in the calculated data for the relativistic Bethe logarithm \cite{korobov2013}.

{\bf 1.} The one-loop vacuum polarization:
\begin{equation}\label{VP}
\Delta E_{vp}^{(7)} =
\frac{\alpha^5}{\pi}
   \biggl[ V_{61}\ln(Z\alpha)^{-2}+V_{60} \biggr]
   \left\langle V_{\delta} \right\rangle
   \approx 2.9\mbox{ kHz}.
\end{equation}
For $S$-states in the hydrogen atom these coefficients are
\[
\left\{
\begin{array}{@{}l}
\displaystyle
V_{61}(nS) = -\frac{2}{15},
\qquad \mbox{\cite{layzer}}
\\[2mm]\displaystyle
V_{60}(nS) = \frac{4}{15}\biggl[
      -\frac{431}{105}+\psi(n+1)-\psi(1)
\\[2mm]\displaystyle\hspace{15mm}
      -\frac{2(n-1)}{n^2}+\frac{1}{28n^2}-\ln{\frac{n}{2}}
   \biggr].
\qquad\mbox{\cite{mohr75,karshenboim97}}
\end{array}
\right.
\]
The coefficient $V_{61}$ does not depend on $n$, the logarithmic contribution is thus proportional to the delta-function.
To estimate the nonlogarithmic contribution in (\ref{VP}), we use the approximate electronic wave function $\psi_e(\mathbf{r}_e)\approx N[\psi_{1s}(\mathbf{r}_1)\!+\!\psi_{1s}(\mathbf{r}_2)]$, where $\psi_{1s}$ is the ground state wave function of the hydrogen atom.
The coefficient $V_{60}$ for the $1S$-state is equal to $-0.63357$.

In fact, the latter term should be calculated numerically for the three-body case. But the contribution itself is of order 1 kHz and this simple approximation provides a good estimate.

{\bf 2.} The Wichman-Kroll contribution \cite{mohr83}:
\begin{equation}
\Delta E_{WK}^{(7)} =
\frac{\alpha^5}{\pi} W_{60}
   \left\langle V_{\delta} \right\rangle
   \approx -0.1\mbox{ kHz}.
\end{equation}
Here $W_{60}(nS) = \frac{19}{45}-\frac{\pi^2}{27}$.

{\bf 3.} The complete two-loop contribution \cite{twoloop}:
\begin{equation}
\begin{array}{@{}l}\displaystyle
\Delta E_{2loop}^{(7)} = \frac{\alpha^5}{\pi}
   \left[
      B_{50}
   \right]
   \left\langle Z_1^2 \delta(\mathbf{r}_1)\!+\!Z_2^2 \delta(\mathbf{r}_2) \right\rangle
\\[2mm]\hspace{30mm}
   \approx 10.1\mbox{ kHz}.
\end{array}
\end{equation}
Here $B_{50}=-21.55447(12)$, this contribution is valid for a bound electron in an arbitrary configuration of few point-like Coulomb sources.

\begin{table}[t]
\begin{center}
\begin{tabular}{@{\hspace{4mm}}l@{\hspace{10mm}}d@{\hspace{-4mm}}}
\hline\hline
$\Delta E_{nr}$        &~1\,522\,150\,208.13   \\
$\Delta E_{\alpha^4}$  &         -50\,320.64   \\
$\Delta E_{\alpha^5}$  &           7\,070.28   \\
$\Delta E_{\alpha^6}$  &              113.11   \\
$\Delta E_{\alpha^7}$  &              -10.46(20)\\
$\Delta E_{\alpha^8}$  &               -0.12(12)\\
\hline\\[-3.5mm]
$\Delta E_{total}$     & 1\,522\,107\,060.3(2)\\
\hline\hline
\end{tabular}
\end{center}
\caption{Summary of contributions to the $(36,34)\to(34,32)$ transition frequency of the $^4\mbox{He}^+\bar{p}$ atom (in MHz).}\label{hep}
\end{table}

{\bf 4.} The three-loop contribution is already negligible. For the hydrogen molecular ion fundamental transition it gives \cite{kinoshita90,eides95,Melnikov00}
\begin{equation}
\Delta E_{3loop}^{(7)} = \frac{\alpha^5}{\pi^2}
   \left[
      0.42
   \right]
   \left\langle Z_1 \delta(\mathbf{r}_1)\!+\!Z_2 \delta(\mathbf{r}_2) \right\rangle
   \approx -60\mbox{ Hz}.
\end{equation}

The above is the complete set of contributions at $m\alpha^7$ order in the nonrecoil limit.

In the next order ($m\alpha^8$) we evaluate only the leading $\ln^3(Z\alpha)^{-2}$ contribution. It represents the second order perturbation with two one-loop self-energy operators ($m\alpha^2(Z\alpha)^6$) \cite{karshenboim93}:
\begin{equation}\label{a82l}
\Delta E_{2loop}^{(8)} = \frac{\alpha^6}{\pi^2}\>\Bigl[ -\frac{8}{27} \Bigr]\ln^3(Z\alpha)^{-2}
   \left\langle V_{\delta} \right\rangle
   \approx 1\mbox{ kHz}.
\end{equation}
Using its value we determine the theoretical uncertainty of yet uncalculated terms in the $m\alpha^8$ order and higher.

\section{Numerical results}

The numerical approach to the two-center problem has been already described in~\cite{korobov2013,tsogbayar2006}; briefly, the following expansion for the electronic wave function is used:
\begin{equation}\label{exp}
\Psi_m(\mathbf{r}_1,\mathbf{r}_2) = e^{im\varphi}r^{|m|}\sum^{\infty}_{i=1}
 C_{i}e^{-\alpha_{i} r_1 - \beta_{i} r_2},
\end{equation}
where $r$ is the distance from the electron to the $z$-axis and $\phi$ the azimuthal angle. For $Z_1=Z_2$ the variational wave function should be symmetrized
\begin{equation}\label{expsym}
\begin{array}{@{}l}\displaystyle
   \Psi_m(\mathbf{r}_1,\mathbf{r}_2) = e^{im\varphi}r^{|m|}\sum^{\infty}_{i=1}
 C_{i}\Bigl(e^{-\alpha_{i} r_{1} - \beta_{i} r_{2}}
\\[2mm]\displaystyle\hspace{30mm}
       \pm e^{-\beta_{i} r_{1} - \alpha_{i} r_{2}} \Bigr),
\end{array}
\end{equation}
where $(+)$ is used to get a {\em gerade} electronic state and $(-)$ is for an {\em ungerade} state, respectively. Parameters $\alpha_{i}$ and $\beta_{i}$ are generated in a quasi-random manner.

We calculated mean values for all operators appearing in Eq.~(\ref{A61_A60}) for the ground ($1s\sigma$) electronic state of the two-center problem, both for $Z_1 = Z_2 = 1$ (H$_2^+$ and HD$^+$ case) and $Z_1 = 2$, $Z_2 = -1$ for the antiprotonic helium. In this way we obtain the coefficients $A_{60}$ and $A_{61}$ (see Fig.~\ref{A6} and \ref{A6_pbarHe}) as well as the other contributions given in Sec. IV B. in the form of effective electronic potential curves.

We then averaged these electronic curves over vibrational wave functions in order to obtain energy corrections for individual states. Adding these new results to previously calculated contributions~\cite{korobov2008,korobov2008b}, one obtains precise theoretical predictions for the frequencies of experimentally relevant transitions (see Tables~\ref{h2p-hdp} and~\ref{hep}). Nonrelativistic energies and leading order corrections were obtained with the CODATA10~\cite{codata2010} recommended values. It is necessary to note that we used improved calculations for the leading order relativistic corrections ($m\alpha^4$) and newly obtained values for the Bethe logarithm \cite{KorZhong2012}, which were the major source of inaccuracy in the leading order radiative corrections ($m\alpha^5$). That allowed to significantly reduce numerical uncertainties in the contributions at these orders.

In the $m\alpha^7$ order the uncertainty on the contribution stems from numerical uncertainty in calculation of the relativistic Bethe logarithm \cite{korobov2013}. The recoil terms are already negligible at order $\alpha^6 (m/M)$, where they contribute about 300 Hz to the fundamental transitions of the hydrogen molecular ion.

The contribution from the finite charge distributions of nuclei deserves special discussion. For the fundamental transition in the H$_2^+$ ion the CODATA10 uncertainty results in 250 Hz uncertainty for the transition energy. If we use instead the charge radius from the muonic hydrogen measurements \cite{muH}, the frequency will move by 3 kHz; ro-vibrational spectroscopy of H$_2^+$ is thus sensitive to the discrepancy between determinations of the proton radius. The CODATA10 uncertainty due to the deuteron rms charge radius for the HD$^+$ fundamental transition is 215 Hz and is so far negligible. In the antiprotonic helium the value of the rms charge radius of the alpha particle is taken from \cite{angeli04} and results in a frequency uncertainty of 7 kHz, while the corresponding uncertainty from the antiproton rms charge radius is more than order of magnitude less, the antiproton--electron interaction being repulsive.

At present most accurate experimental results are available for the HD$^+$ molecular ion and for the antiprotonic helium. In Table \ref{comparison} we compare our new theoretical results with the best experimental ones.  Agreement is excellent in all cases except for the $v=0 \to v=1$ transition in HD$^+$ where the discrepancy is $2.6\,\sigma_{exp}$.

\section{Conclusion}

We have completed the calculation of the $\alpha^7$-order one-loop self-energy correction in two-center systems, and used these results to obtain new predictions of experimentally relevant ro-vibrational transition frequencies for the three-body molecular type systems. The theoretical uncertainty has been improved by about one order of magnitude to reach a level of 0.03 ppb in molecular hydrogen ions (resp. 0.13 ppb in the antiprotonic helium). The achieved accuracy already allows for improved determination of the proton- and antiproton-to-electron mass ratios \cite{codata2010}, and may still be improved further as discussed above. Particularly, as a first step we intend to improve the relativistic Bethe logarithm calculations using the asymptotic expansions for $P_{rc}(k)$, and $P_{nq}(k)$ functions presented in the Appendix. That may result in reducing uncertainty in the one-loop self energy contribution by a factor of three and reduce theoretical relative uncertainty for vibrational transitions to $10^{-11}$.

\section{Acknowledgements}

V.I.K. acknowledges support of the Russian Foundation for Basic Research under Grant No. 12-02-00417. This work was supported by \'Ecole Normale Sup\'erieure, which is gratefully acknowledged.

\begin{table}[t]
\begin{center}
\begin{tabular}{@{\hspace{1mm}}l@{\hspace{3mm}}l@{\hspace{3mm}}l@{\hspace{1mm}}}
\hline\hline
\vrule height 10.5pt width 0pt depth 3.5pt
 & \mbox{~~~experiment} & \mbox{~~~~~theory} \\
\hline
\vrule height 10.5pt width 0pt depth 3.5pt
 $\mbox{HD}^+(v,L)$ \\
$(0,2)\to(4,3)$~\cite{koelemeij2007} & 214\,978\,560.6(5)  & 214\,978\,560.948(8)  \\
$(0,0)\to(1,1)$~\cite{bressel2012} & \hspace{1.7mm}58\,605\,052.00(6) & \hspace{1.7mm}58\,605\,052.156(2) \\
$(0,0)\to(0,1)$~\cite{Shen12}      & \hspace{9mm} ---                 & \hspace{3.3mm}1\,314\,925.7523(1) \\
$(0,2)\to(8,3)$~\cite{koelemeij2012}&\hspace{9mm} ---                 & 383\,407\,177.150(15) \\
\hline
\vrule height 10.5pt width 0pt depth 3.5pt
 $^4\mbox{He}^+\bar{p}\,(n,L)$\hspace{2.9mm}\cite{hori2011} \\
$(36,34)\!\to\!(34,32)$& 1\,522\,107\,062(4)  & 1\,522\,107\,060.3(2)  \\
$(33,32)\!\to\!(31,30)$& 2\,145\,054\,858(5)  & 2\,145\,054\,858.1(2)  \\
 $^3\mbox{He}^+\bar{p}\,(n,L)$ \\
$(35,33)\!\to\!(33,31)$& 1\,553\,643\,100(7) & 1\,553\,643\,102.4(3) \\
\hline\hline
\end{tabular}
\end{center}
\caption{Comparison with most accurate experimental measurements of transition frequencies for HD$^+$ and antiprotonic helium (in MHz). The two transitions \cite{koelemeij2012,Shen12} are currently studied experimentally and for convenience of future comparison we present our theoretical values for these transitions.}\label{comparison}
\end{table}

\appendix

\section{Asymptotic expansion of $P_{rc}^{(1)}(k)$, $P_{rc}^{(2)}(k)$, and $P_{nq}(k)$}

Here we present without proof the results for the asymptotic expansion of the functions defined in Eqs.~(\ref{Prc1})--(\ref{Pnq}), which appear in the integrand of (\ref{BLau}). Of particular importance is the term of order $1/k^2$, which contributes to the $\alpha^6\ln{\alpha}$ part of Eq.~(\ref{a7au}). It is finite and the $\pi Z^3\left\langle\delta(\mathbf{r})\right\rangle$ "counterterm" is determined by the choice of regularization of the divergent operators. As in \cite{korobov2013} we separate $P_{rc}^{(1)}$ into two parts:
\begin{subequations}
\begin{equation}
\begin{array}{@{}l}\displaystyle
P^{(1a)}_{rc}(k) =
   \Bigl\langle
      \mathbf{p}\left(E_0\!-\!H\!-\!k\right)^{-1}
      \Bigl(H_B\!-\!\left\langle H_B \right\rangle\Bigr)
\\[2mm]\displaystyle\hspace{35mm}
      \left(E_0\!-\!H\!-\!k\right)^{-1}\mathbf{p}
   \Bigr\rangle
\end{array}
\end{equation}
\begin{equation}
\begin{array}{@{}l}\displaystyle
P^{(1b)}_{rc}(k) =
   2\Bigl\langle
      H_BQ(E_0\!-\!H)^{-1}Q\mathbf{p}
\\[2mm]\displaystyle\hspace{35mm}
      \left(E_0\!-\!H\!-\!k\right)^{-1}\mathbf{p}
   \Bigr\rangle \, .
\end{array}
\end{equation}
\end{subequations}

In the following expressions, the second line gives the numerical values of the asymptotic expansion coefficients for the 1S state of the hydrogen atom.
\begin{widetext}
\begin{equation}
\begin{array}{@{}l}\displaystyle
P_{rc}^{(1a)}(k) = -\frac{\sqrt{2}}{k^{3/2}}\,\pi Z^2\left\langle\delta(\mathbf{r})\right\rangle
   \!+\!\frac{8\ln{2}\!-\!13}{2k^2}\,\pi Z^3\left\langle\delta(\mathbf{r})\right\rangle
   +\frac{1}{16k^2}\left\langle (\nabla^4V) \right\rangle_{\rm fin}
   -\frac{1}{k^2}\left\langle
      \left(H_B\!-\!\left\langle H_B \right\rangle\right) \boldsymbol{\nabla}^2
   \right\rangle_{\rm fin}
   +\dots
\\[3mm]\displaystyle\hspace{14mm}
 = -\frac{\sqrt{2}}{k^{3/2}}+\frac{8\ln{2}+1}{2k^2}+\dots
\end{array}
\end{equation}
\vspace{-2mm}
\begin{equation}
\begin{array}{@{}l}\displaystyle
P_{rc}^{(1b)}(k) = \frac{2}{k}\left\langle
      \left( H_B\!-\!\left\langle H_B \right\rangle\right)
      (E_0\!-\!H)^{-1}\boldsymbol{\nabla}^2
   \right\rangle
   +\frac{2\sqrt{2}}{k^{3/2}}\,\pi Z^2\left\langle\delta(\mathbf{r})\right\rangle
   +\frac{\ln{k}}{k^2}\,\pi Z^3\left\langle\delta(\mathbf{r})\right\rangle
   +\frac{5\ln{2}\!-\!1}{k^2}\,\pi Z^3\left\langle\delta(\mathbf{r})\right\rangle
\\[3mm]\displaystyle\hspace{20mm}
   +\frac{1}{k^2}\left\langle
      \left( H_B\!-\!\left\langle H_B \right\rangle\right)
      (E_0\!-\!H)^{-1}(\boldsymbol{\nabla}^2V)
   \right\rangle_{\rm fin}
   +\frac{1}{k^2}\left\langle
      \left( H_B\!-\!\left\langle H_B \right\rangle\right)
      \boldsymbol{\nabla}^2
   \right\rangle_{\rm fin}
   +\dots
\\[3mm]\displaystyle\hspace{14mm}
 = -\frac{2}{k}
   +\frac{2\sqrt{2}}{k^{3/2}}+\frac{\ln{k}}{k^2}
   +\frac{3\ln{2}-6}{k^2}
   +\dots
\end{array}
\end{equation}
\begin{equation}
\begin{array}{@{}l}\displaystyle
P_{rc}^{(2)}(k) = \frac{\left\langle \nabla^4 \right\rangle}{k}
   -\frac{8\sqrt{2}}{k^{3/2}}\,\pi Z^2\left\langle\delta(\mathbf{r})\right\rangle
   +\frac{4\ln{k}}{k^2}\,\pi Z^3\left\langle\delta(\mathbf{r})\right\rangle
\\[3mm]\displaystyle\hspace{20mm}
   -\frac{12\ln{2}+4}{k^2}\,\pi Z^3\left\langle\delta(\mathbf{r})\right\rangle
   -\frac{1}{2k^2}\left\langle(\boldsymbol{\nabla}^2V)p^2\right\rangle_{\rm fin}
   -\frac{1}{k^2}\left\langle(\boldsymbol{\nabla}V)^2\right\rangle_{\rm fin}
   +\dots
\\[3mm]\displaystyle\hspace{13.5mm}
 = \frac{5}{k}-\frac{8\sqrt{2}}{k^{3/2}}
   +\frac{4\ln{k}}{k^2}
   -\frac{20\ln{2}\!-\!18}{k^2}
   +\dots
\end{array}
\end{equation}
\vspace{-2mm}
\begin{equation}
\begin{array}{@{}l}
\displaystyle
P_{nq}(k) = -\frac{1}{2}
      \bigl\langle
         \boldsymbol{\nabla}^2
      \bigr\rangle
   -\frac{1}{5k}
      \bigl\langle
         \boldsymbol{\nabla}^4
      \bigr\rangle
   -\frac{1}{2k}\left\langle (\nabla^2V) \right\rangle
   +\frac{8\sqrt{2}}{k^{3/2}}\>
      \pi Z^2\left\langle \delta(\mathbf{r}) \right\rangle
   -\frac{8\ln{k}}{k^2}\>
      \pi Z^3\left\langle \delta(\mathbf{r}) \right\rangle
\\[4mm]\hspace{20mm}\displaystyle
   +\frac{40\ln{2}+76}{5k^2}\>\pi Z^3\left\langle \delta(\mathbf{r}) \right\rangle
   +\frac{2}{k^2}\left\langle(\boldsymbol{\nabla}V)^2\right\rangle_{\rm fin}
   +\frac{3}{40k^2}\left\langle(\nabla^4V)\right\rangle_{\rm fin}
   +\frac{1}{2k^2}\left\langle(\nabla^2V)p^2\right\rangle_{\rm fin}
\\[5mm]\hspace{10.5mm}\displaystyle
 = \frac{1}{2}-\frac{3}{k}+\frac{8\sqrt{2}}{k^{3/2}}
   -\frac{8\ln{k}}{k^2}
   +\frac{24\ln{2}\!-\!10}{k^2}+\dots
\end{array}
\end{equation}
\end{widetext}
The sum of these terms makes up the result given in Eq.~(\ref{asymp}).

It is worth noting here that this is the main point where we differ from the approach used in \cite{jentschura2005}, where the formal expansion over $1/k$ has been used:
\begin{equation}
\frac{1}{E_0\!-\!H\!-\!k} = -\frac{1}{k}-\frac{E_0\!-\!H}{k^2}-\frac{(E_0-H)^2}{k^3}+\dots,
\end{equation}
which gives divergent matrix elements for individual $S$ states in the hydrogen-like atom, but still the "normalized difference" $\Delta_n$ is finite. This formalism is enough to get a complete result for arbitrary states in the hydrogen atom [see Eq.~(3.43) and Table 1 of Ref.~\cite{jentschura2005}], but not suitable for our generalization. So we took another way, which is to derive an appropriate approximation to the $\psi_1$ function
\begin{equation}\label{psi1}
\psi_1 = (E_0\!-\!H\!-\!k)^{-1}\>i\mathbf{p}\>\psi_0,
\end{equation}
where $\psi_0$ is a stationary solution of the Schr\"odinger equation. $\psi_1$ is a regular function at small $r$  thus providing finite expectation values for the operators appearing at $1/k^2$ order. This procedure is very similar to what was used to obtain the asymptotic expansion for the nonrelativistic Bethe logarithm in \cite{korobov12}.

\end{document}